\documentclass[prd,reprint,nofootinbib,showpacs,superscriptaddress]{revtex4-1}

\usepackage{graphicx} 
\usepackage{hyperref}
\usepackage{amsfonts}
\usepackage{amsmath,amssymb}
\usepackage{bm} 
\usepackage{color}
\usepackage{epstopdf} 
\usepackage{epsfig}
\usepackage{subfig}

\def\be{\begin{equation}} 
\def\ee{\end{equation}}  
\def\bea{\begin{eqnarray}}
\def\eea{\end{eqnarray}}
\def\ba{\begin{array}}
\def\ea{\end{array}}
\def\bem{\begin{multline}} 
\def\eem{\end{multline}}  

\newcommand{\lb}{\,\,\, b}
\newcommand{\llb}{\,\,\,\,\,\,\,\,\,\,\,\, b}

\begin{document}

\title{Enhancement of Critical Temperature of a Striped Holographic Superconductor}

\author{Suman Ganguli}
\email{sgangul3@tennessee.edu}
\affiliation{Department of Physics and Astronomy, The University of Tennessee, Knoxville, Tennessee 37996-1200}
\author{Jimmy A. Hutasoit}
\email{jimmy.hutasoit@mail.wvu.edu} 
\affiliation{Department of Physics, West Virginia University, Morgantown, West Virginia 26506}

\author{George Siopsis}
\email{siopsis@tennessee.edu}
\affiliation{Department of Physics and Astronomy, The University of Tennessee, Knoxville, Tennessee 37996-1200}

\date{\today} 

\begin{abstract}
We study the interplay between the stripe order and the superconducting order in a strongly coupled striped superconductor using gauge/gravity duality. In particular, we study the effects of inhomogeneity introduced by the stripe order on the superconducting transition temperature beyond the mean field level by including the effects of backreaction onto the spacetime geometry in the dual gravitational picture. We find that inhomogeneity \emph{enhances} the critical temperature relative to its value for the uniform system.
\end{abstract}

\maketitle


\section{Introduction}

One of the differences between conventional superconductors and high temperature superconductors is that the normal states of the conventional superconductors are well described by Fermi liquid, whose only (weak coupling) instability is to superconductivity. By contrast, the normal states high temperature superconductors, such as cuprates and iron pnictides, are highly correlated and thus, exhibit other low temperature orders which interact strongly with superconductivity. One of the prominent orders is the unidirectional charge density wave ``stripe" order \cite{Hoffman:2002,Howald:2003,Vershinin:2004,Wise:2008}.

It is therefore important to understand the nature of the interplay between superconductivity and the stripe order in the presence of strong correlation. In this article, we study the effects of the stripe order on the superconducting transition temperature of the strongly correlated superconductor using holography or gauge/gravity correspondence. In gauge/gravity duality, the strongly coupled condensed matter systems are mapped to a weakly coupled Einstein-Maxwell-scalar theory on black hole spacetimes with negative cosmological constant, or the so-called anti de-Sitter (AdS) black holes. Just like the normal states of high temperature superconductors, AdS black holes feature numerous types of instability that lead to the formation of scalar 'hair' \cite{Gubser:2008vn,Hartnoll:2008ys} (which corresponds to superconductivity), striped phases \cite{Nakamura:2009fk,Aperis:2010kx,Donos:2011uq} and nematic phases \cite{Edalati:2012zr}. Ultimately, we would like to study the system where both the superconducting order and the stripe order emerge dynamically, however, since our focus in this article is the effects of the stripe order on the critical temperature, we will follow Refs. \onlinecite{Flauger:2011vn, Hutasoit:2012fk} where the inhomogeneity is introduced via a modulated chemical potential, with wavenumber $Q$. These articles studied the holographic striped superconductor by neglecting the backreaction of the electromagnetic field on the spacetime geometry, which in field theory language corresponds to the mean field treatment of the system. To improve upon their results, we would like to consider the effects of fluctuations on the strongly coupled striped superconductor by studying the backreacted spacetimes.

Our result shares similar qualitative features as the result of Ref. \cite{Martin:2005fk}, which study the effects of local inhomogeneity on the critical temperature within the framework of BCS theory. This similarity is of course not surprising, as one can reproduce the weak coupling result from the strong coupling result by simply taking the limit in which the scaling dimension of the superconducting order becomes the scaling dimension of a free scalar field \cite{Hutasoit:2012fk}.

We find that the critical temperature for the formation of the scalar hair is maximum in the case of AdS Schwarzschild black hole, which is the solution to the Einstein equation when backreaction is neglected \cite{Flauger:2011vn, Hutasoit:2012fk}. This means that the superconducting transition temperature obtained by including fluctuations is lower than that obtained in the case where fluctuations are neglected. In the field theory side, this can also be understood using a Ginzburg-Landau type argument: introducing fluctuations costs us free energy and thus lowering the critical temperature. Usually, the corrections due to fluctuations are small enough that the qualitative behavior of $T_c$ as a function of other parameters remains the same. However, when the system is inhomogeneous due to the presence of stripe order, the effects of fluctuations at the regime where the inverse fluctuation length scale is smaller than the wavenumber $Q$ are rather drastic. Starting at $Q=0$, when we increase $Q$, we see that $T_c$ exhibits a steep jump as we turn on the modulation and after reaching a maximum, it monotonously decreases as $Q$ increases and asymptotes to a constant value at $Q \to \infty$. As the critical temperature at finite $Q$ is larger than the values at $Q=0$ and $Q \to \infty$, which correspond to the homogenous limits, we find that the critical temperature is enhanced by the presence of stripe order.

\section{Set-up}
\label{sec:2}

We are interested in studying a strongly coupled striped superconductor using the gauge/gravity duality. To this end, consider a $U(1)$ gauge potential $A^a$ and a scalar field $\psi$  charged under this potential living in a spacetime with negative cosmological constant $\Lambda = -3/L^2$. The scalar field is dual to the scalar order parameter of the superconductor, i.e., the condensate, while the $U(1)$ gauge field is dual to the four-current in the condensed matter system. For simplicity, we shall adopt units in which $L=1$, $16\pi G =1$.

To study the strong coupling regime of the superconductor, we only need to study the gravity theory at the classical level. In particular, we are interested in finding solutions to the classical equations of motion whose boundary values are related to the parameters of the superconductor.

The action for this system is
\be\label{ActionEMS}
S =\int d^4x\sqrt{-g}\left[ R + 6 - \frac{1}{4} F^2 - |D_a\psi|^2 - m^2|\psi|^2 \right]\,,
\ee
where, $D_a=\partial_a-iqA_a$, $F_{ab}=\partial_{a} A_{b}-\partial_{b} A_{a}$ and $a,b\in\{ t,r,x,y\}$. Here

The field equations consist of the Einstein equations,
\be\label{EinsteinEqn1}
R_{ab} - \frac12g_{ab} R - 3g_{ab} =\frac{1}{2} T_{ab}\, , \ee
where the stress-energy tensor is
\bea
T_{ab} &=& F_{ac} F_{b}^{\,\,\,c} - \frac14g_{ab} F^{cd} F_{cd} \nonumber\\
 &+ &  D_a\psi (D_b\psi)^* + (a \leftrightarrow b ) - g_{ab} \left[ |D_a\psi|^2 + m^2 |\psi|^2 \right] \,,\nonumber\\
\eea
the Maxwell equations,
\be\label{MaxwellEqn1} 
\frac{1}{\sqrt{-g}}\partial_{b}(\sqrt{-g}F^{ab}) = J^{a}\,,
\ee
where the $U(1)$ current is
\be J^{a}=-i[\psi^* D^{a}\psi- \mathrm{c.c.} ] \,, \ee
and the Klein-Gordon equation for the scalar field,
\be\label{ScalarEqn1}
-\frac{1}{\sqrt{-g}}D_{a}(\sqrt{-g}g^{ab}D_{b}\psi)+m^2 \psi=0\,.
\ee

\section{Above the critical temperature} 
\label{sec:3}

Above the critical temperature $T_c $, the scalar field vanishes ($ \psi =0$) and thus,
the Einstein-Maxwell equations simplify to
%
\be\label{EinsteinEquation2} 
R^a_{\lb} + 3 \delta^a_{\lb}  = \frac{1}{2} \, T^a_{\lb}  \, , \ \ \ \
\partial_{b}(\sqrt{-g}F^{ab}) = 0\,,
\ee
where
\be\label{Def Stress Tensor 1} 
T^a_{\lb}  =   F^{a c} F_{bc} - \frac{1}{4} \delta^a_{\lb}   F^{cd} F_{cd} \,.
\ee
We are interested in finding a static black hole solution of flat conformal boundary which is sourced by an external modulated chemical potential $\mu (\vec x)$, where $\mu$ is a given spatially dependent function.
Let $(x,y)$ be the Cartesian spatial coordinates of the two-dimensional conformal boundary. We concentrate on the case in which $\mu$ only depends on one of the coordinates, which is chosen to be $x$. Expanding in Fourier cosine modes, we write
\be \mu (x) = \mu \sum_{n=0}^\infty \delta_n \cos nQx \,, \ \ \ \ \sum_{n=0}^\infty \delta_n =1. \ee
For definiteness, we consider the case in which only the first couple of modes are non-vanishing, setting
\be\label{eq10} \delta_0 = 1- \delta \,, \ \ \ \ \delta_1 = \delta \, , \ \ \ \ \delta_n = 0 \ \ (n\ge 2). \ee
To find a solution to the Einstein-Maxwell equations \eqref{EinsteinEquation2}, consider the metric {\em ansatz}
\be\label{MetricAnsatz}
ds^2 = -r^2 e^{-\alpha} dt^2+ e^{\alpha} \frac{dr^2}{r^2} + r^2 e^{-\beta} \left[ e^{-\gamma} dx^2 + e^{\gamma} dy^2  \right]\,,
\ee
where $\alpha$, $\beta$, and $\gamma$ are functions of $(r,x)$. The boundary is at $r\to\infty$ and the horizon is at $r=r_+$, where $r_+$ is an arbitrary parameter. For a flat conformal boundary, we require $\alpha$, $\beta$, $\gamma \to 0$, as $r\to\infty$, and in fact, we find $\alpha \sim \mathcal{O} (r^{-3})$ while  $\beta$ and $\gamma \sim \mathcal{O} (r^{-4})$.

For the $U(1)$ potential, we fix the gauge such that $A_r=A_x=A_y=0$ and $A_t = A_t(r,x)$ with $A_t=0$ at the horizon (for a finite norm, $A_aA^a < \infty$), whereas at the boundary,
\be A_t (r,x) \Big|_{r\to\infty} = \mu (x)\, .\ee
We shall solve  the Einstein-Maxwell equations \eqref{EinsteinEquation2} perturbatively in $ \mu^2 $. This expansion is valid for large black holes (or small chemical potential), or more precisely for
\be \mu \lesssim r_+\, . \ee
Expanding in the small dimensionless parameter $\mu / r_+$, we have
\bea\label{ExpansionAg}
A_t &=& A^{(0)}_t + \left(\frac{\mu}{r_+}\right)^2 \, A^{(1)}_t + \dots \nonumber\\ 
\alpha &=& \alpha^{(0)} + \left(\frac{\mu}{r_+}\right)^2 \, \alpha^{(1)} +\dots \nonumber\\
\beta &=& \beta^{(0)} + \left(\frac{\mu}{r_+}\right)^2 \, \beta^{(1)} +\dots \nonumber\\
\gamma &=& \gamma^{(0)} + \left(\frac{\mu}{r_+}\right)^2 \, \gamma^{(1)} +\dots  
\eea
and consequently, the expansions of the metric, Ricci tensor and $U(1)$ field strength and stress-energy tensor, respectively,
\bea g_{ab} &=& g_{ab}^{(0)} + \left(\frac{\mu}{r_+}\right)^2 g_{ab}^{(1)} + \dots \nonumber\\
R_{ab} &=& R_{ab}^{(0)} + \left(\frac{\mu}{r_+}\right)^2 R_{ab}^{(1)} + \dots \nonumber\\
F_{ab} &=&  \left(\frac{\mu}{r_+}\right)^2 F_{ab}^{(0)} + \dots \nonumber\\
T_{ab} &=& \left(\frac{\mu}{r_+}\right)^2 \mathcal{T}_{ab}^{(0)} + \dots \eea
At zeroth order, the Einstein-Maxwell equations read
\be\label{Expansion of Einstein's Equation}
R^{(0) \,a}_{\llb} + 3 \delta^a_{\lb} =0 \,, \ \ \ \ \partial_b \left( \sqrt{-g^{(0)}} F^{(0)ab} \right) = 0\, . \ee
The Einstein equations decouple and are solved by the AdS Schwarzschild black hole
\be\label{eqh} e^{-\alpha^{(0)}} \equiv h = 1 - \left( \frac{r_+}{r} \right)^3 \,, \ \ \ \ \beta^{(0)} = \gamma^{(0)} = 0\,. \ee
To solve the Maxwell equations, it is convenient to introduce the coordinate
\be z = \frac{r_+}{r}\,, \ee
so that the boundary is at $z=0$ and the horizon at $z=1$. Expanding the $U(1)$ potential in Fourier modes
\be A_t^{(0)} = \mu \sum_{n=1}^\infty \delta_n \mathcal{A}_n (z) \cos nQx\,, \ee
we deduce the mode equations
\be \mathcal{A}_n''(z) - \frac{n^2Q^2}{r_+^2 h(z)} \mathcal{A}_n(z) = 0 \ , \ee
to be solved together with the boundary conditions $\mathcal{A}_n(0) =1$, $\mathcal{A}_n(1)=0$. Here, $h(z) = 1-z^3$ (Eq.\ \eqref{eqh}) and $'$ denotes a derivative with respect to $z$. For $n=0$, we obtain
\be \mathcal{A}_0 (z) = 1-z \,.\ee
For $n\ge 1$, a good analytic approximation to the solution is given by
\be \mathcal{A}_n (z) \approx \frac{\sinh \left[\frac{n \, Q}{r_+} (1-z)\right]}{\sinh \frac{nQ}{r_+}} \,.\ee
With the choice of boundary conditions \eqref{eq10}, the lowest-order stress-energy tensor $\mathcal{T}_{ab}^{(0)}$ has modes with $n\le 2$ due to the fact that it is quadratic in the $U(1)$ potential. The same should be true for the first-order corrections to the metric.

Explicitly, the non-vanishing components of the zeroth-order electromagnetic stress-energy tensor are
\bea \mathcal{T}_{\,\,\,\,\,\,\,\,\,\,\,\, t}^{(0)\, t} &=& -\mathcal{T}_{\,\,\,\,\,\,\,\,\,\,\,\, y}^{(0)\, y} = - \frac{z^4}{4} \left[ \frac{\mathcal{E}_x^2}{h} + \mathcal{E}_z^2 \right] \,,\nonumber\\
\mathcal{T}_{\,\,\,\,\,\,\,\,\,\,\,\, z}^{(0)\, z} &=& -\mathcal{T}_{\,\,\,\,\,\,\,\,\,\,\,\, x}^{(0)\, x} = \frac{z^4}{4} \left[ \frac{\mathcal{E}_x^2}{h} - \mathcal{E}_z^2 \right]\,, \nonumber\\
\mathcal{T}_{\,\,\,\,\,\,\,\,\,\,\,\, z}^{(0)\, x} &=& \frac{1}{h} \mathcal{T}_{\,\,\,\,\,\,\,\,\,\,\,\, x}^{(0)\, z} = - \frac{z^4}{2 \, h} \mathcal{E}_x \mathcal{E}_z \,,
\eea
given in terms of the components of the electric field
\bea\label{Electric Fields} \mathcal{E}_x &=& \frac{\delta Q}{r_+} \mathcal{A}_1 \sin Qx \,, \nonumber\\
\mathcal{E}_z &=& (1-\delta) \mathcal{A}_0'(z) + \delta \mathcal{A}_1'(z) \cos Qx \,.\eea
To solve the Einstein equations at first order,
\be R_{\llb}^{(1)\, a} = \mathcal{T}_{\llb}^{(0)\, a} \,,\ee
we set
\bea\label{eqabc} \alpha^{(1)} &=& \alpha_0^{(1)}(z) + \alpha_1^{(1)} (z) \cos Qx + \alpha_2^{(1)} (z) \cos 2Qx \,,\nonumber\\
\beta^{(1)} &=& \beta_0^{(1)}(z) + \beta_1^{(1)} (z) \cos Qx + \beta_2^{(1)} (z) \cos 2Qx \,, \nonumber\\
\gamma^{(1)} &=& \gamma_0^{(1)}(z) + \gamma_1^{(1)} (z) \cos Qx + \gamma_2^{(1)} (z) \cos 2Qx \,. \nonumber\\
\eea
We obtain five non-vanishing components for each set of functions $\{ \alpha_i^{(1)} , \beta_i^{(1)} , \gamma_i^{(1)} \}$, where $i=0,1,2$.
Of the five equations, only three are independent and can be solved analytically for the three corresponding metric functions.
After some algebra, we obtain the following system of equations for the modes of the metric functions.

For the Fourier zero modes, we obtain
\bea\label{Metric Eqn 03}  
\alpha_{0}^{(1)\prime} -\left(\frac{3}{z}-\frac{h'}{h}\right)\alpha_{0}^{(1)}-\frac{z}{2}\left(\frac{4}{z}-\frac{h'}{h}\right)\gamma_{0}^{(1)\prime} & & \nonumber\\
- \frac{z^3 \left(\frac{Q^2}{r_+^2} \delta^2 \, \mathcal{A}_1^2- h \left(2 (1-\delta )^2 \mathcal{A}_0'^2+\delta ^2 \mathcal{A}_1'^2\right)\right)}{8 h^2} &=& 0 \,,\nonumber\\
\beta_{0}^{(1)\prime\prime}-\left(\frac{2}{z}-\frac{h'}{h}\right) \beta_{0}^{(1)\prime}-\frac{ Q^2 z^2 \delta ^2 \mathcal{A}_1^2}{4 r_+^2 h^2} &=& 0\,,
\nonumber\\
\gamma_{0}^{(1)\prime\prime}+\frac{ Q^2 z^2 \delta ^2 \mathcal{A}_1^2}{4 r_+^2 h^2} &=& 0\,.
\nonumber\\
\eea
We solve these equations by requiring that the functions be regular at the horizon ($z=1$) and vanish sufficiently fast at the boundary ($z=0$).
We obtain
\be 
\gamma_{0}^{(1)} (z) = -  \frac{Q^2 \delta^2}{4\,r_+^2} \int_0^z \, dz' \int_0^{z'} dz''  \ \frac{(z'')^2 \,\mathcal{A}_1^2}{h^2}\,,
\ee
\be 
\beta_{0}^{(1)} (z) = -  \frac{Q^2 \delta^2}{4\,r_+^2} \int_0^z dz' \, \frac{(z')^2}{h} \int_{z'}^1 dz'' \, \frac{\mathcal{A}_1^2}{h} \, ,
\ee
\be\label{theta0zz} 
\alpha_{0}^{(1)}(z)=\frac{z^3 }{8h} \int_z^1 \overline{\alpha}_0^{(1)}(z')  \, dz' \, ,\ee
where
\be \overline\alpha_0^{(1)}(z) = 2 (1-\delta )^2 {\mathcal{A}_0'}^2+\delta ^2 {\mathcal{A}_1'}^2  -\frac{Q^2 }{r_+^2} \delta ^2 \frac{\mathcal{A}_1^2}{h}- \gamma_0^{(1)\prime} \frac{4 h-z h'}{z^3}\,.
\ee
For the Fourier first modes, we obtain
\bea \label{Metric Eqn 13}
\alpha_{1}^{(1)\prime}-\left(\frac{3}{z}-\frac{\frac{Q^2}{r_+^2}  \, z+2h'}{2h}\right) \alpha_{1}^{(1)} \qquad \qquad & & \nonumber\\
-\frac{ z}{2} \left(\frac{4}{z}-\frac{h'}{ h} \right) \gamma_{1}^{(1)\prime}
+ \frac{ Q^2 z (\beta_{1}^{(1)}-\gamma_{1}^{(1)})}{2 r_+^2 h} & & \nonumber\\
+\frac{z^3 \delta (1-\delta ) \mathcal{A}_0'\mathcal{A}_1'}{2 h} &=& 0 \,,
\nonumber\\
\beta_{1}^{(1)\prime\prime}-\left(\frac{2}{z}-\frac{h'}{h}\right) \beta_{1}^{(1)\prime} &=& 0\,,
\nonumber\\
\gamma_{1}^{(1)\prime\prime}-\frac{Q^2 }{r_+^2 \, h} \alpha_1^{(1)} &=& 0\,.
\eea
The second equation readily yields
\be 
\beta_{1}^{(1)} (z) =0\,.
\ee
By eliminating $\alpha_1^{(1)}$ between the other two equations, we obtain a third order differential equation for $\gamma_1^{(1)}$. Then the possible behavior of $\gamma_1^{(1)}$ at the horizon is found to be a linear combination of $1-z$, $(1-z)\ln (1-z)$, and $(1-z)^{1+Q^2/ 6r_+^2 }$. We fix the three integration constants by demanding $\gamma_1^{(1)} (0) = 0$, $\gamma_1^{(1)\prime} (0) =0$, and $\gamma_1^{(1)\prime\prime} \lesssim \mathcal{O} (1/(1-z))$ at the horizon ($z=1$). The second boundary condition, together with Eqs.\ \eqref{Metric Eqn 13}, ensure $\alpha_1^{(1)} \sim z^3$ at the boundary. The third boundary condition is necessary for the existence of a well-defined temperature (surface gravity), resulting in $\alpha_1^{(1)} (1) =0$, on account of the third equation in \eqref{Metric Eqn 13}. 

Finally, for the Fourier second modes, we obtain
\bea \label{Metric Eqn 23}
\alpha_{2}^{(1)\prime}-\left(\frac{3 }{z }-\frac{ 2\frac{Q^2}{r_+^2} z+h'}{ h}\right) \alpha_{2}^{(1)} \qquad \qquad & & \nonumber\\
-\frac{z}{2}\left(\frac{4}{z}- \frac{h'}{h}\right) \gamma_{2}^{(1)\prime} + \frac{2 Q^2 z }{r_+^2 h} (\beta_{2}^{(1)}-\gamma_{2}^{(1)}) & & \nonumber\\
+\frac{\delta ^2  z^3 \left(\frac{Q^2}{r_+^2} \mathcal{A}_1^2+ h {\mathcal{A}_1'}^2\right)}{8 h^2} &=& 0\,,
\nonumber\\
\beta_{2}^{(1)\prime\prime}-\left(\frac{2}{z}-\frac{h'}{h}\right) \beta_{2}^{(1)\prime}+\frac{\delta ^2 Q^2 z^2 \mathcal{A}_1^2}{4 r_+^2h^2} &=& 0\,,
\nonumber\\
\gamma_{2}^{(1)\prime\prime}-\frac{Q^2\left(16 \, h \, \alpha_{2}^{(1)}+\delta^2 z^2 \mathcal{A}_1^2\right)}{4 r_+^2 h^2} &=& 0\,.
\eea
The second equation yields
\be 
\beta_{2}^{(1)} (z) = \frac{Q^2 \, \delta^2}{4r_+^2}\int_0^z dz' \, \frac{(z')^2}{h}\int_{z'}^1 dz''  \, \frac{\mathcal{A}_1^2}{h} \,.
\ee
We note that $\beta_2^{(1)} = - \beta_0^{(1)}$.

Eliminating $\alpha_2^{(1)}$ between the other two equations, we obtain, as before, a third-order differential equation for $\gamma_2^{(1)}$, from which we deduce the possible near horizon behavior, $1-z$, $(1-z)\ln (1-z)$, and $(1-z)^{1+2Q^2 /3r_+^2}$. As before, we fix the three integration constants by demanding $\gamma_2^{(1)} (0) = 0$, $\gamma_2^{(1)\prime} (0) =0$, and $\gamma_2^{(1)\prime\prime} \lesssim \mathcal{O} (1/(1-z))$ at the horizon ($z=1$). The second boundary condition, together with Eqs.\ \eqref{Metric Eqn 23}, ensure $\alpha_2^{(1)} \sim z^3$ at the boundary. The third boundary condition is necessary for the existence of a well-defined temperature (surface gravity), resulting in $\alpha_2^{(1)} (1) =0$, on account of the third equation in \eqref{Metric Eqn 23}.

\begin{figure}[ht]
\begin{center}  
\includegraphics[scale=0.5]{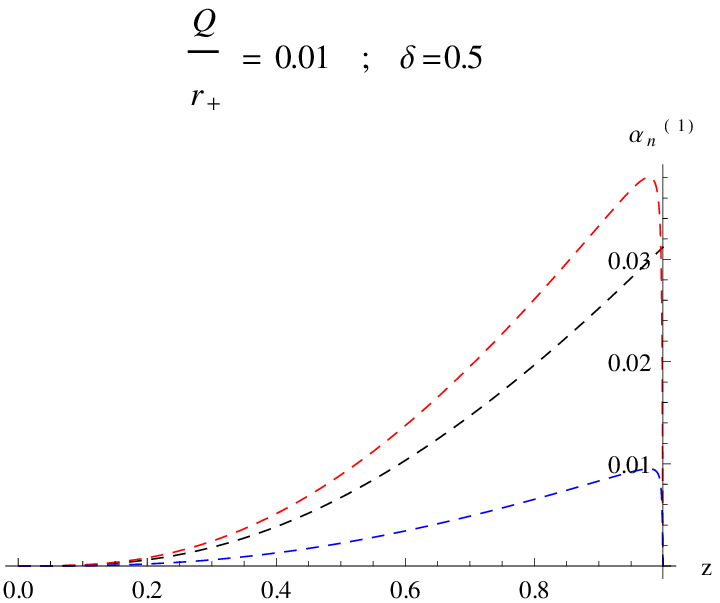}     
\hspace{0.2cm}\includegraphics[scale=0.5]{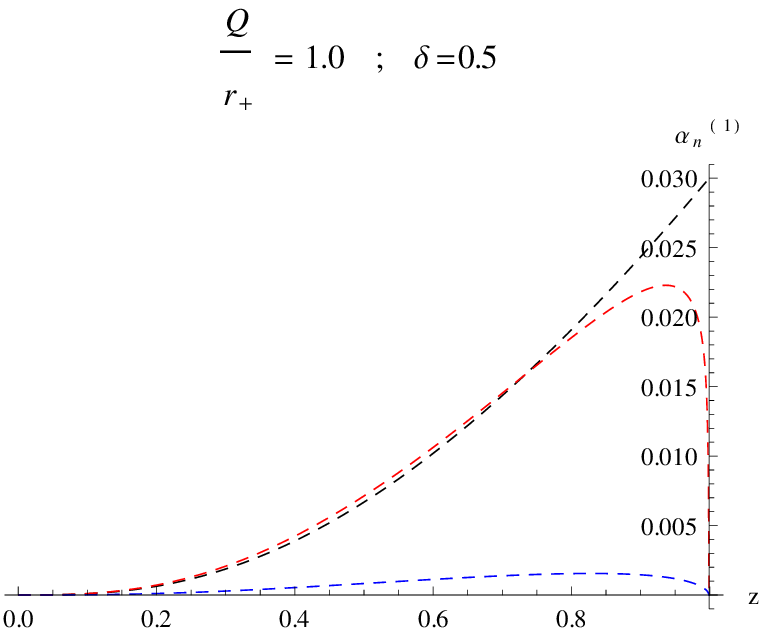}  
\includegraphics[scale=0.5]{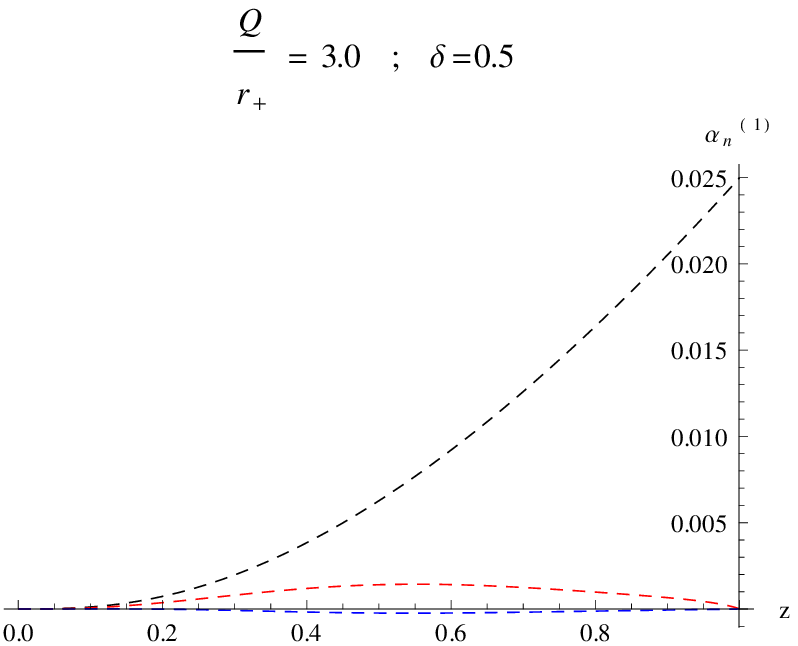}     
\hspace{0.2cm}\includegraphics[scale=0.5]{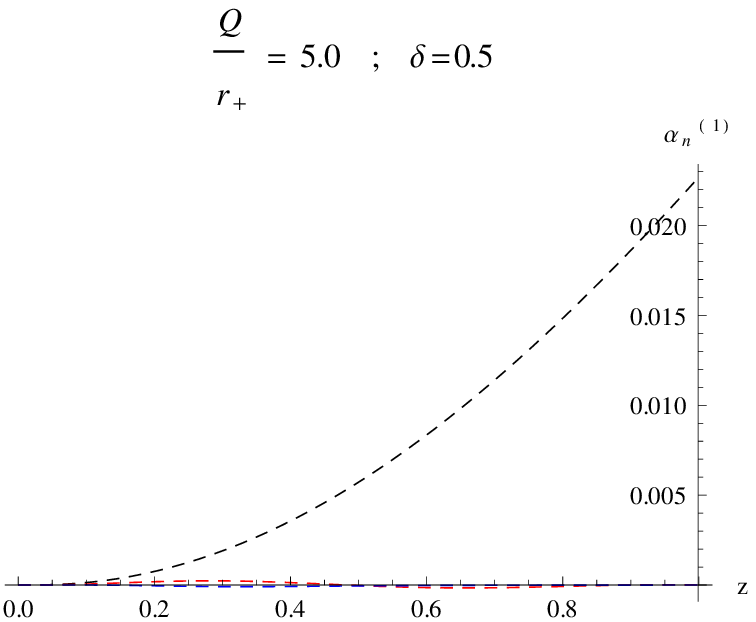}  
\end{center}  
\caption{$\alpha_0^{(1)}$ (black), $\alpha_1^{(1)}$ (red) and $\alpha_2^{(1)}$ (blue) for $\delta=0.5$, and $Q/r_+ = 0.1$, $1$, $3$ and $5$. \label{fig: alpha}}  
\end{figure}

\begin{figure}[ht]
\begin{center} 
\includegraphics[scale=0.8]{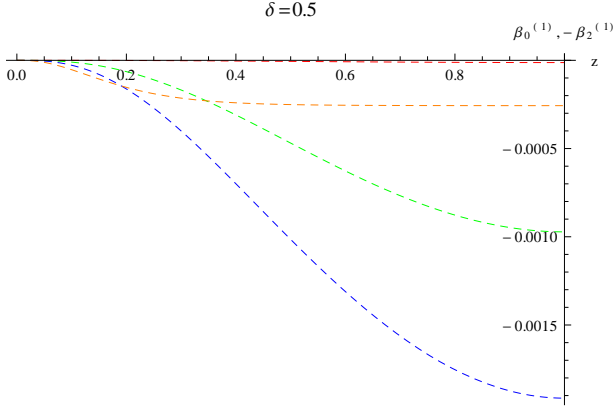} 
\end{center}  
\caption{$ \beta_0^{(1)} =-\beta_2^{(1)} $ for $ \delta=0.5 $, and $ QL^2/r_+ = $ $0.1$ (red), $1.0$ (green), $2.0$ (blue) and $8.0 $ (orange). \label{fig: beta}}   
\end{figure}

\begin{figure}[ht]
\begin{center} 
\includegraphics[scale=0.8]{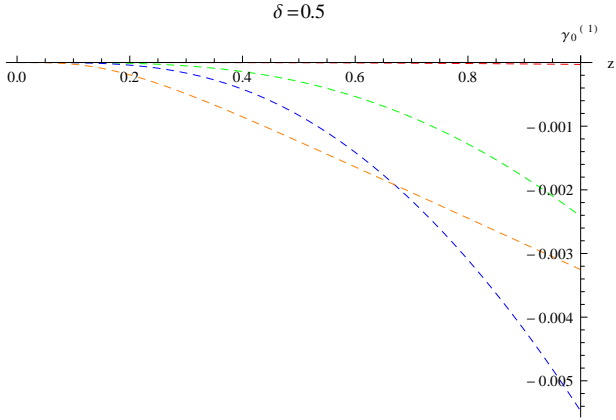}       
\end{center}  
\caption{ $ \gamma_0^{(1)} $ for $ \delta=0.5 $, and $ Q/r_+ = $ $0.1$ (red), $1.0$ (green), $2.0$ (blue) and $8.0 $ (orange). \label{fig: gamma0}}    
\end{figure}

\begin{figure}[ht] 
\begin{center} 
\includegraphics[scale=0.8]{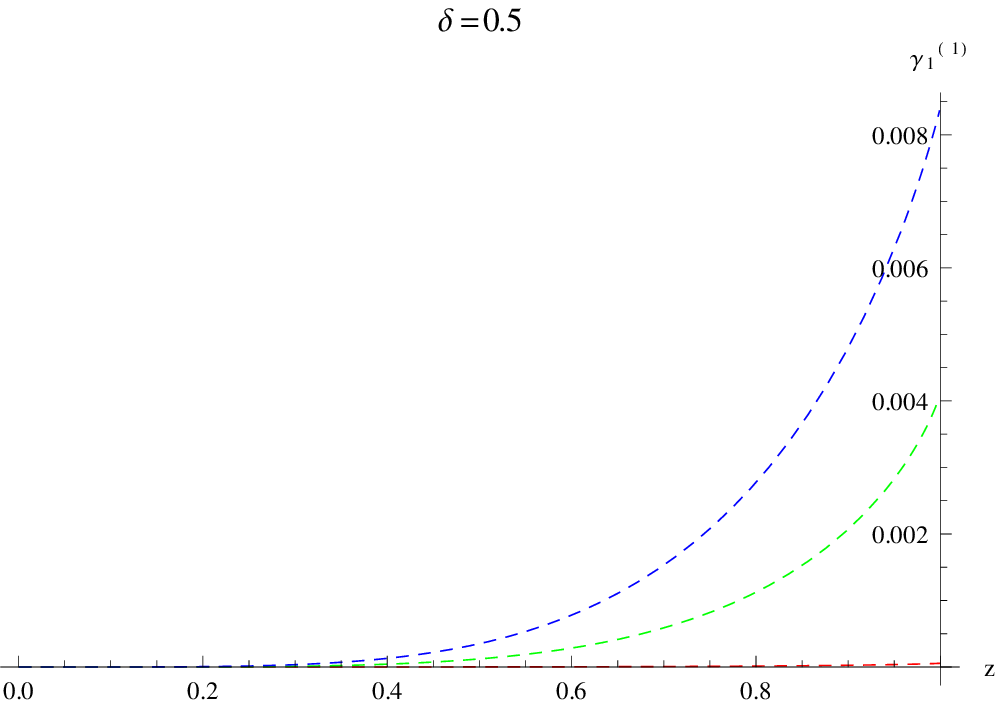}
 \end{center} 
\caption{ $ \gamma_1^{(1)} $ for $ \delta=0.5 $, and $ Q/r_+ = $ $0.1$ (red), $1.0$ (green) and $2.0$ (blue). \label{fig: gamma1 }}   
\end{figure}

\begin{figure}[ht]
\begin{center} 
\includegraphics[scale=0.8]{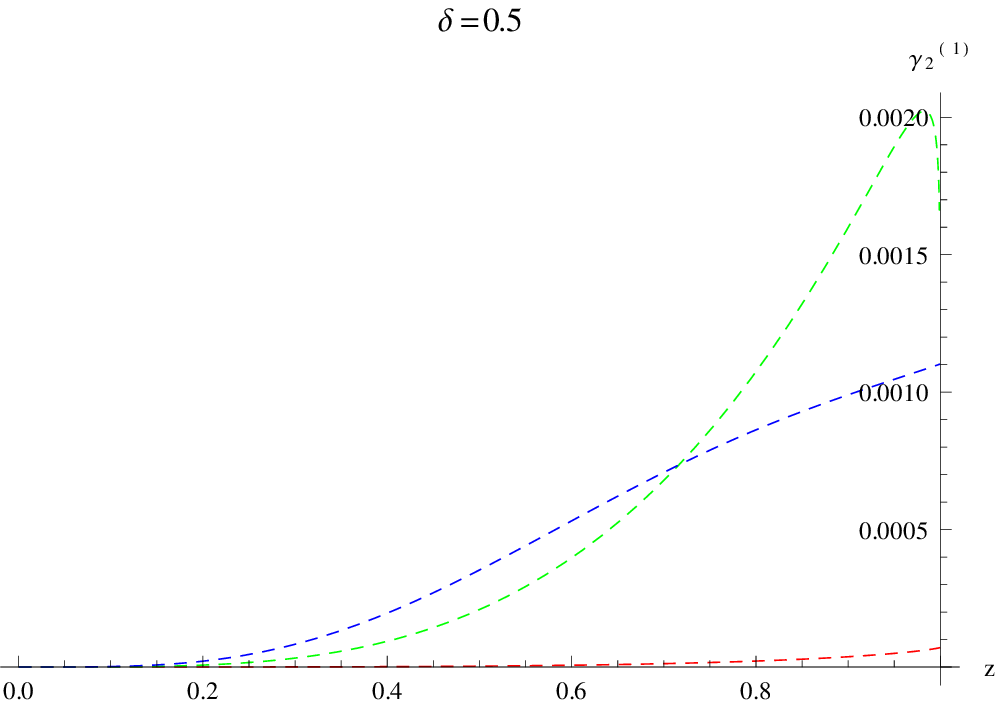}
\end{center}  
 \caption{ $ \gamma_2^{(1)} $ for $ \delta=0.5 $, and $ Q/r_+ = 0.1$ (red), $1.0$ (green) and $2.0$ (blue).
 \label{fig: gamma2 } }
\end{figure}

The equations for the various modes can be solved numerically subject to the boundary conditions outlined above. We have plotted $\alpha_n^{(1)}$ ($n=0,1,2$) in Fig. \ref{fig: alpha} for representative values of $Q$, whereas $\beta_n^{(1)}$ ($n=0,2$; it vanishes for $n=1$) is plotted in Fig. \ref{fig: beta}, and $\gamma_n^{(1)}$ is plotted in Figs. \ref{fig: gamma0}, \ref{fig: gamma1 }, and \ref{fig: gamma2 }, for $n=0,1$ and $2$, respectively.


Note that, the $ \beta_{n}^{(1)} $ and $\gamma_{n}^{(1)}  $ components of metric perturbations are sourced by $ x $-component of electric field \eqref{Electric Fields}, which vanishes at both small and large $ Q $.
The functions $ \beta_{n}^{(1)} $s and $\gamma_{n}^{(1)}  $s depend on $ Q $ via two terms: a direct proportionality factor $ Q^2 $ and area under the functions $ \mathcal{A}_1^2/h $ or $ z^2 \mathcal{A}_1^2/h^2 $.
The first factor vanishes at $ Q \to 0 $, while the integrals vanish at $ Q \to \infty $, due to $\mathcal{A}_1 \sim \frac{Q}{\sinh Q}\, (1-z) $ near the horizon.
Consequently, $\beta_n^{(1)}$ and $\gamma_n^{(1)}$ ($n=0,1,2$) are very small in both limits $Q\ll r_+$ and $Q\gg r_+$. These functions are more significant in the intermediate range $2 < Q/r_+ < 3$ and decay rapidly on both sides, but even when they reach their maximum, they remain well below unity (see Figs. \ref{fig: beta}--\ref{fig: gamma2 }). Thus, their contribution to physical quantities is negligible in the entire range of $Q$.

Next, we discuss the behavior of $\alpha_n^{(1)}$ ($n=0,1,2$) which are physically important because they determine the temperature. Indeed, the Hawking temperature at first perturbative order is
\be\label{eqTH} T = \frac{3r_+}{4\pi } \left[ 1 - \frac{\mu^2}{r_+^2} \alpha^{(1)} (1) \right] \,. \ee
Since $\alpha_n^{(1)} (1) =0$ for $n\ge 1$, we have
\bea \alpha^{(1)} (1) &=& \alpha_0^{(1)} (1) = \frac{\overline{\alpha}_0 (1)}{24} \nonumber\\
&=&
\left. \frac{2 (1-\delta )^2 +\delta ^2 {\mathcal{A}_1'}^2 - 3\gamma_0^{(1)\prime} }{24} \right|_{z=1} \, .\eea
We can calculate these functions analytically in the two important limits: $Q\to 0$ and $Q\to\infty$.

In the limit $Q\to 0$, we obtain the analytic expressions
\bea  
\alpha_{0}^{(1)} &=& \left( (1-\delta )^2 +\frac{\delta ^2}{2} \right) \frac{z^3}{4\left(1+z+z^2\right)} + \mathcal{O} \left( \frac{Q^2}{r_+^2} \right) \,,
\nonumber\\
\alpha_{1}^{(1)} &=& \frac{(1-\delta ) \,\delta \,z^3}{2\left(1+z+z^2\right)} (1-z)^{Q^2/6r_+^2} + \mathcal{O} \left( \frac{Q^2}{r_+^2} \right) \,,
\nonumber\\
\alpha_{2}^{(1)} &=& \frac{\delta ^2\, z^3}{8\left(1+z+z^2\right)} (1-z)^{2Q^2/3r_+^2} + \mathcal{O} \left( \frac{Q^2}{r_+^2} \right) \,.
\eea
At $Q=0$ (or equivalently, $\delta =0$), we recover the exact Reissner-Nordstr\"om solution representing the homogeneous system
\bea\label{gtt at Q=0}  
e^{-\alpha} &=& e^{-\alpha^{(0)}} \left( 1+\frac{\mu^2}{r_+^2} \, \alpha^{(1)} \right) \nonumber\\
&=& 1- \left( 1+ \frac{ \mu^2   }{4r_+^2} \right) z^3 + \frac{\mu^2}{4r_+^2} z^4 ~,
\eea
where we used $\alpha^{(1)} = \alpha_0^{(1)} + \alpha_1^{(1)} + \alpha_2^{(1)}$ from Eq.\ \eqref{eqabc} with $Q=0$.

We recover the Schwarzschild solution, which is the solution in the probe limit, in the limit $\mu\to 0$. As we increase $\mu$, we move further away from the probe limit and the effects of back reaction to the metric become more pronounced. We reach extremality at $\mu /r_+ = 2\sqrt{3}$, which gives the upper bound for $\mu$.

It should be noted that the convergence to the homogeneous system is not uniform. At the horizon, $\alpha_n^{(1)} (1) \to 0$ for $n=1,2$, and therefore $\alpha^{(1)}$  does not converge to its homogeneous counterpart. In other words, the limits $Q\to 0$ and $z\to 1$ do not commute.
It follows that there is a discontinuity in the temperature which depends on the behavior of $\alpha_n^{(1)} $ at the horizon. From Eq.\ \eqref{eqTH} in the limit  $Q\to 0$, we obtain
\be\label{eqTHq} T \approx \frac{3r_+}{4\pi } \left[ 1 - \frac{\mu^2 \left( (1-\delta )^2 + \delta^2/2 \right)}{12r_+^2} \right] \,, \ee
which is valid for small $Q$.
Comparing this result with the homogeneous case, which is recovered by setting $\delta =0$, we obtain an \emph{enhancement} in temperature upon turning on modulation
\be\label{eqenh} \frac{\Delta T}{T} =  \frac{T}{T_{\delta =0}} -1\approx \frac{\mu^2}{12r_+^2} \delta \left( 2- \frac{3\delta}{2} \right)\,, \ee
with a maximum enhancement for $\delta = \frac{2}{3}$.
The change in temperature is discontinuous, but this is an artifact of keeping only the first order in perturbation theory. This change in the temperature is expected to become smooth (yet remain steep) as higher orders in the perturbative expansion are included.

On the other hand, in  the $ Q \gg r_+ $ regime, the contribution of $\mathcal{A}_1$ becomes exponentially small, and all functions except $ \alpha_{0}^{(1)} $ become negligible. In this regime, we have
\be\label{gtt at Q=inf}  
\alpha_{0}^{(1)} \approx  \frac{ (1-\delta )^2 \, z^3}{4\left(1+z+z^2\right)} \,.
\ee
So in the $ Q \to \infty $ limit, we recover another exact Reissner-Nordstr\"om solution, albeit with less charge density,
\be 
e^{-\alpha} \approx 1 - \left( 1+ \frac{ \mu^2  (1-\delta )^2  }{4r_+^2} \right) z^3 + \frac{\mu^2 (1-\delta)^2}{4r_+^2} z^4 \,.
\ee
This coincides with the homogeneous solution \eqref{gtt at Q=0} if $\delta =0$, as expected.

We then deduce the temperature for large $Q$ to be given by 
\be\label{eqTHQ} T \approx \frac{3r_+}{4\pi } \left[ 1 - \frac{\mu^2 (1-\delta )^2}{12r_+^2} \right] \,. \ee

\section{The Critical Temperature} 
\label{sec:4}
The Klein-Gordon equation for a static scalar field $\psi(z,x)$ of mass $m$ and charge $q$ reads
\begin{multline}\label{Scalar PDE}  
\sum_{i=z,x} \frac{1}{\sqrt{-g}}\, \partial_i \left( \sqrt{-g} g^{ii} \partial_i \psi \right) +  \left( q^2 \, A_t^2  - m^2 \right) \psi =0\,.
\end{multline} 
The mass is related to the conformal dimension $\Delta$ of the superconducting order parameter by
\be 
m^2 = \Delta \, \left( \Delta - 3 \right)\,.
\ee
To solve the wave equation,
we expand $ \psi $ in a Taylor series around $x=0$
\be\label{Def Psi} \psi(z,x) = \frac{\langle \mathcal{O}_\Delta \rangle}{\sqrt{2}}   z^\Delta
\left[ F_0 (z) + x^2 F_1 (z) + \dots \right] \,.\ee
The series is expected to converge rapidly at the two ends ($Q\to 0$ and $Q\to\infty$), where the $x$-dependence of $\psi$ is mild. For intermediate values of $Q$, the $x$-dependence is more significant, but the zero mode $F_0$ still dominates. In this regime, the error in the numerical analysis can be reduced to the desired accuracy by including higher modes of the Fourier expansion. We shall leave this to future work, as our focus here is the temperature enhancement at small $Q$, where the calculation involves slowly varying functions of $x$.

We obtain the leading order wave equation
\begin{multline}\label{Scalar 0-th mode equation} 
F_0'' + \left[\frac{2 (\Delta -1)}{z}+\frac{\overline{h}'}{\overline{h}}\right]F_0'\\
+\frac{\Delta  }{z^2 \overline{h}} \left[\, (\Delta -3) (\overline{h}-1)+z \, \overline{h}' \, \right] \, F_0 +  \frac{q^2 \, (\overline{A}_t^{(0)})^2 }{r_+^2 \, \overline{h}^2} F_0 =0 \,,
\end{multline} 
where
\be 
\overline{h} \equiv e^{-\alpha} \Big|_{x=0} = h \, \left[ 1- \frac{\mu^2}{r_+^2} \left( \alpha_{0}^{(1)} +\alpha_{1}^{(1)}+\alpha_{2}^{(1)} \right)   \right] \Big|_{x=0} \,,
\ee 
and $\overline{A}_t^{(0)} (z) \equiv A_t^{(0)} (z,0)$.

To solve this wave equation, it is convenient to fix the parameters $ \lbrace \Delta, \delta,q,Q/r_+ \rbrace  $ and find the eigenvalue for
\be \lambda = \frac{q \mu}{ r_+} \,.\ee
The critical temperature is then determined by Eq.\ (\ref{Scalar 0-th mode equation}) and by the value of $ T/r_+ $, which is given by      
 \eqref{eqTH}. We obtain
\be \frac{T_c}{q\mu} = \frac{3}{4\pi} \left[ \frac{1}{\lambda} - \frac{\lambda}{q^2} \alpha^{(1)} (1) \right] \,.\ee
At small $Q$, we obtain from Eq.\ \eqref{eqenh} the enhancement in critical temperature
\be\label{eqDTc} \frac{\Delta T_c}{T_c} \approx \frac{\lambda^2}{12 q^2} \delta \left( 2 - \frac{3\delta}{2} \right) \,,\ee
which vanishes at the probe limit ($q\to\infty$) and becomes significant away from it.
However, we need to be cautious in taking the small $q$ limit, as this is only a first-order result, which is $\mathcal{O} (1/q^2)$.

\begin{figure}[ht]
\begin{center} 
\includegraphics[scale=0.9]{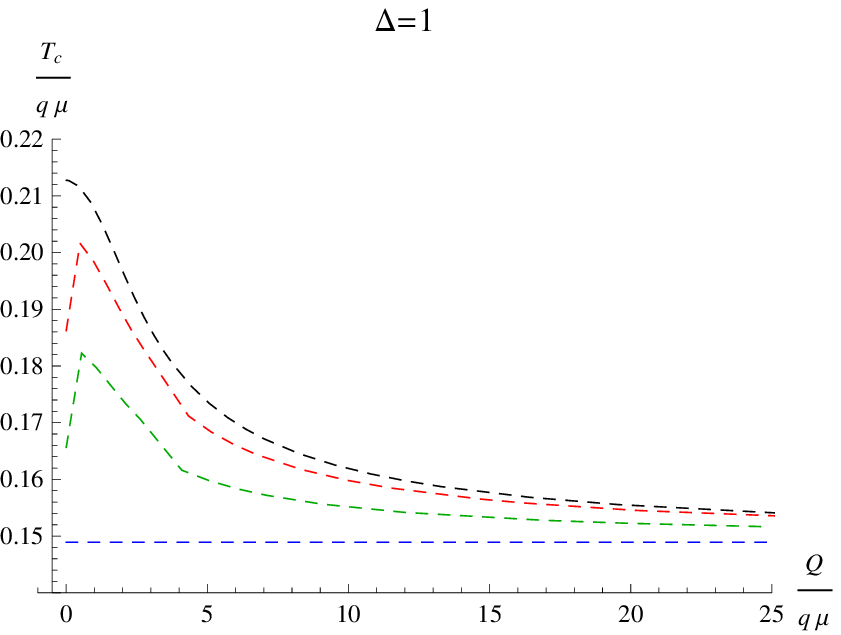}
\end{center} 
\caption{From top to bottom: $T_c $ \emph{vs.} $ Q $ for $ \Delta=1 $ and $(\delta , q^2) = (0.3,\infty)$, $(0.2, 0.411)$ and $(0.1, 0.190)$. Parameters are chosen so that curves asymptote to $T_c/(q\mu) = 0.149$ as $Q\to\infty$.\label{fig: TcvsQ Delta 1}}     
\end{figure}

\begin{figure}[ht]
\begin{center} 
\includegraphics[scale=0.9]{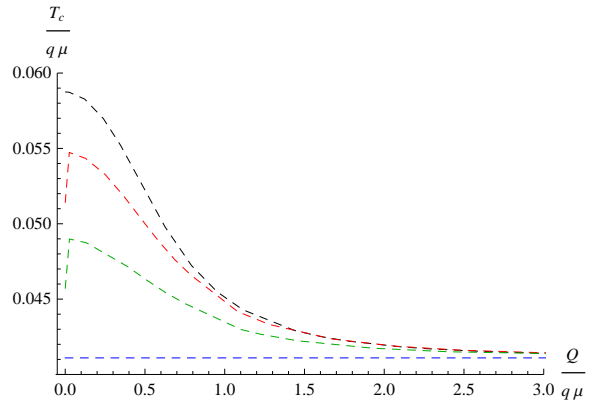}
\end{center} 
\caption{From top to bottom: $ T_c $ \emph{vs.} $ Q $ for $ \Delta=2 $ and $(\delta , q^2) = (0.3,\infty)$, $(0.2, 7.92)$ and $(0.1, 4.22)$. Parameters are chosen so that curves asymptote to $T_c/(q\mu) = 0.041$ as $Q\to\infty$. \label{fig: TcvsQ Delta 2}}     
\end{figure}

The wave equation is solved numerically subject to the boundary conditions $F_0 \sim z^\Delta$ at the boundary and the demand of regularity at the horizon ($F_0(1) < \infty$).
The results are shown in Figs. \ref{fig: TcvsQ Delta 1} and \ref{fig: TcvsQ Delta 2}, for $\Delta =1$ and $2$, respectively.
In each case, we have chosen the other parameters so that the curves asymptote to the same temperature as $Q\to \infty$. We note that all curves exhibit a jump at $Q=0^+$, showing the enhancement of the critical temperature once modulation is switched on, in agreement with our analytic result \eqref{eqDTc}. As $Q$ increases, the critical temperature decreases monotonically. The jump vanishes in the probe limit which is obtained for $\mu =0$ (Schwarzschild black hole). For any given $Q$, the critical temperature attains its maximum value at this limit. Put differently, back reaction to the metric lowers the critical temperature. Correspondingly, in the dual boundary system, quantum fluctuations result in a reduction in the critical temperature for a given modulation vector $Q$.

\section{Summary and Outlook}  
\label{sec:5}  

In this article, we have studied the effect of inhomogeneity on the superconducting transition temperature of the strongly coupled striped superconductor beyond the mean field level, by including backreaction of the electromagnetic field on the geometry of spacetime in the dual gravitational picture. We found that as we turn on the modulation, the critical temperature exhibits a steep jump. After that, as we increase $Q$, the critical temperature decreases until it reaches the asymptotic value. In other words, we found an enhancement of the critical temperature due to inhomogeneity that comes from the stripe order.

The discontinuous jump we see here is an artifact of only keeping the zero mode of the scalar field in the calculation and we expect that as we include the higher modes, this jump will become smooth but yet steep. It will be interesting to study whether the maximum of $T_c$ corresponds to the value of $Q$ being the inverse of superconducting correlation length scale as is seen in the BCS result.

\acknowledgments{We would like to thank Ivar Martin for insightful discussions. J.\ H.\ is supported by West Virginia University start-up funds. The work of S.\ G.\ and G.\ S.\ is supported in part by the Department of Energy under grant DE-FG05-91ER40627.} 

\bibliographystyle{kp}
\bibliography{References} 
\end{document}